\documentclass[12pt,a4paper]{article}
\usepackage{latexsym,amsmath,amssymb}
\date{}
\numberwithin{equation}{section}

\begin{document}
\title{Supersymmetric Distributions, Hilbert Spaces of Supersymmetric Functions and Quantum Fields \\}
\author{Florin Constantinescu\\ Fachbereich Mathematik \\ Johann Wolfgang Goethe-Universit\"at Frankfurt\\ Robert-Mayer-Strasse 10\\ D 60054
Frankfurt am Main \\ Germany } \maketitle

\begin{abstract}
The recently investigated Hilbert-Krein and other positivity structures of the superspace are considered in the
framework of superdistributions. These tools are applied to problems raised by the rigorous supersymmetric quantum
field theory.
\end{abstract}

\section{Introduction}

In spite of their formidable success on many areas of physics, the path integral methods are in most cases formal.
In the case of supersymmetry, even methods of moderate rigor like canonical quantization are not generally worked
out. Certainly supersymmetric quantum fields have to be operators but the (positive definite) Hilbert space, in
which they are supposed to act, and the associated domain problems, are outside formal methods. Whereas it is easy
to see that the Hamiltonian in supersymmetry is positive \cite{WB,We}, the argument does not give further
information on the nature and realization of the Hilbert space behind.\\ In \cite{C1} we have proved that the $N=1
$ superspace shows an intrinsec Hilbert-Krein structure realized on supersymmetric functions (not fields). Taking
up this finding, in this note, we answer general questions on supersymmetric distributions and Hilbert spaces
together with their first application to what it was called axiomatic quantum field theory \cite{SW}. If we
pretend rigor they have to get definite answers before pursuing further developments. \\ In this paper we
concentrate on the "massive" case in quantum field theory. The "massles" case requires a little more effort but
does not require new tools. The main point of the massless case is a natural restriction of the space of
supersymmetric functions to a subspace of it \cite{C1}. This restriction is necessary in order to maintain the
intrinsec Hilbert-Krein structure and on the other hand remembers rigorous methods of the canonical quantization
of gauge theories (Gupta-Bleuler, Nielsen-Lautrup, Kugo-Ojima etc., see \cite{C1}). \\ Concluding, we may say that
the new point of the rigorous supersymmetric quantum field theory (and of this paper too) rests on the
Hilbert-Krein structure which is present already in the massive case, contrary to the non-supersymmetric quantum
fields where a Hilbert-Krein structure turns out to be necessary only in the massless case \cite{St}.

\section{Test functions and distributions for supersymmetry}

Our supersymmetric (test) functions and distributions and the Hilbert spaces related to them are not as general as
usually used in the mathematical literature on supersymmetry and supermannifolds \cite{R,DW,FP}. In our work we
are guided by direct applications to the supersymmetric quantum field theory \cite{WB}. Let $x$ be in the
4-dimensional Minkowski space with the signature $(-1,1,1,1)$ and let $\theta ,\bar \theta $ be two-component
Grassmann variables associated to $x$. We use the notations and conventions in \cite{WB}. They coincide with those
of \cite{S} up to the Pauli $\sigma^0 $ which in \cite{WB} and this paper is equal the 2x2-matrix minus one (it is
one in \cite{S}). In means that in this paper $\bar \sigma $ is defined to be $\bar \sigma =(\sigma^0 ,-\sigma^1
,-\sigma^2 ,-\sigma^3 )=(-1,-\sigma^1 ,-\sigma^2 ,-\sigma^3 )$ where $\sigma =(\sigma^l ),l=1,2,3 $ are the Pauli
matrices. The most general function of $z=(x,\theta ,\bar \theta )$ is of the form

\begin{gather}\nonumber
X(z)=X(x,\theta ,\bar \theta )= \\ \nonumber =f(x)+\theta \varphi (x) +\bar \theta \bar \chi (x) +\theta
^2m(x)+\bar \theta^2n(x)+ \\ +\theta \sigma^l\bar \theta v_l(x)+\theta^2\bar \theta \bar \lambda(x)+\bar
\theta^2\theta \psi (x)+ \theta^2 \bar \theta^2d(x)
\end{gather}
where the coefficients of $\theta ,\bar \theta $ are functions of $x$. For the vector component $v$ we can write
equivalently
\[\theta \sigma^l\bar \theta v_l=\theta ^{\alpha }\bar \theta ^{\dot \alpha }v_{\alpha \dot \alpha } \]
where \[v_{\alpha \dot \alpha }=\sigma _{\alpha \dot \alpha }^l v_l,v^l=-\frac{1}{2}\bar \sigma^{l\dot \alpha
\alpha }v_{\alpha \dot
  \alpha } \]
In principle $x$ is a number but at a certain stage of the computations related to supersymmetry we will be forced
to admit that $x$ is not only a number (base) but also contains an even element of the Grassmann algebra. In this
case we perform the Taylor expansion retaining for $x$ only the base. Functions of several variables $z_1 ,z_2
,...$ have also expansions in $\theta_1 ,\bar \theta_1 ,\theta_2 ,\bar \theta_2 ,... $ of the same type. In what
follows we use a mixed van der Waerden calculus in which $\theta ,\bar \theta $ are Grassmann variables commuting
with the component of the "spinor" coefficients of $X$. The rules of this calculus are \cite{C1}:

\begin{gather}
\psi \chi =\psi^\alpha \chi_\alpha =-\psi_\alpha \chi^\alpha =-\chi^\alpha \psi_\alpha =-\chi \psi \\ \bar \psi
\bar \chi =\bar \psi_{\dot \alpha }\bar \chi^{\dot \alpha }=-\bar \psi^{\dot \alpha }\bar \chi_{\dot \alpha
}=-\bar \chi_{\dot \alpha }\bar \psi^{\dot \alpha }=-\bar \chi \bar \psi \\ \overline {\chi \psi }=\overline
{\chi^\alpha \psi_\alpha }=\bar \chi^{\dot \alpha }\bar \psi_{\dot \alpha }=\bar \psi_{\dot \alpha }\bar
\chi^{\dot \alpha }=\bar \psi \bar \chi =-\bar \chi \bar \psi =-\overline {\psi \chi }
\end{gather}
together with

\begin{gather}(\theta \phi )(\theta \psi )=\frac{1}{2}(\phi \psi )\theta^2 \\
(\bar \theta \bar \phi )(\bar \theta \bar \psi )=\frac{1}{2}(\bar \phi \bar \psi )\bar \theta^2 \\ \chi \sigma^n
\bar \psi =\bar \psi \bar \sigma^n \chi \\ \overline {\chi \sigma^n \bar \psi }=-\bar \chi \bar \sigma^n
\psi=-\psi \sigma^n \bar \chi \\ \overline {\bar \chi \bar \sigma^n \psi }=-\chi \sigma^n \bar \psi =-\bar \psi
\bar \sigma^n \chi
\end{gather}
They differ from the usual ones in which the Grassmann variables together with the components of the spinor fields
are all anticommuting \cite{WB,S}. In this later case, following the terminology in physics, $X$ is said to be a
field (but not yet the operator quantum field to be defined later). As a consequence we make difference between
(supersymmetric) functions (or distributions), fields and quantum fields to be defined later. Note that some
combined relations as for instance
\[ \overline {\chi \sigma^n \bar \psi }=-\bar \chi \bar \sigma^n \psi \]
remain unchanged with respect to the mixed or usual van der Waerden calculus. Taking $\bar \theta $ to be the
(Grassmann) conjugate of $\theta $, $\bar \theta_{\dot \alpha }=\overline {\theta_{\alpha}} $, with
$\overline{\theta_{1\alpha }\theta_{2\beta }\ldots \theta_{n\gamma}}=\bar \theta_{n\dot \gamma}\ldots \bar
\theta_{2\dot \beta } \bar \theta_{1\dot \alpha }$ we have for functions

\begin{gather} \nonumber
\bar X=\bar X(x,\theta ,\bar \theta )=\\ \nonumber =\bar f(x)-\theta \chi (x) -\bar \theta \bar \varphi (x)
+\theta ^2 \bar n(x)+\bar \theta^2 \bar m(x)+ \\ +\theta \sigma^l \bar \theta \bar v_l(x)-\theta^2 \bar \theta
\bar \psi(x)-\bar \theta^2\theta \lambda (x)+ \theta^2 \bar \theta^2 \bar d(x)
\end{gather}
where the bar represents either the complex, the Grassmann or both complex und Grassmann conjugations (the one or
another meaning of the bar will be clear from the context).  This relation is different from the usual one in
which Grassmann variables anticommute with spinor components (i.e. the case of fields) which is

\begin{gather} \nonumber
\bar X=\bar X(x,\theta ,\bar \theta )=\\ \nonumber =\bar f(x)+\theta \chi (x) +\bar \theta \bar \varphi (x)
+\theta ^2 \bar n(x)+\bar \theta^2 \bar m(x)+ \\ +\theta \sigma^l \bar \theta \bar v_l(x)+\theta^2 \bar \theta
\bar \psi(x)+\bar \theta^2\theta \lambda (x)+ \theta^2 \bar \theta^2 \bar d(x)
\end{gather}
by minus signs of the odd terms. The relation $\overline {\overline X }=X $ holds invariably. \\

Now we pass to superdistributions. The cheapest way, which we adopt in the present paper, is to define them again
by (2.1) where the coefficients of $\theta ,\bar \theta $ are now distributions instead of functions. The
definition of functions and distributions can be extended to several variables. The standard order of the
Grassmann variables is supposed to be $ \theta_1 ,\bar \theta_1 ; \theta_2 ,\bar \theta_2 ;... $. For convenience
we denote by $S(R^{4n}\times G )$ the set of functions in $n$ variables $z_1 ,z_2 ,\ldots ,z_n $ with coefficients
in $S(R^{4n})$ where $G$ stays for Grassmann. Some problems appear at the point we want to pass to the duality
functional i.e. if we want to look at superdistributions as linear continuous functionals on spaces of
supersymmetric test functions (supersymmetric duality). Certainly the duality functional has to extend integration
including Grassmann variables (Berezin integration). In one variable this means

\begin{gather}
(\mathcal{X},X)=\mathcal{X}(X)=\int d^8 z\mathcal{X}X
\end{gather}
where $X$ is the test function and $\mathcal{X}$ the would be distribution. The notations and conventions
regarding the Berezin integration are as in \cite{S}. Note that we have fixed the order in which $\mathcal{X}$ and
$X$ appear in the integral (2.12). One way to cope with duality would be to introduce a locally convex topology on
the space of test functions. This can be done by using seminorms (norms) suggested in \cite{R,FP}. Another way is
the following. The function $X$ can be identified with the vector valued function $X$ with components $(f,\varphi
,\bar \chi ,m,n,v,\bar \lambda ,\psi ,d)$. We assume the coefficient functions to be indefinitely differentiable
and may give to the linear space of vector valued functions $X$ the locally convex topology common for the
Schwartz spaces $\mathcal{D}$ (or $ S $). By usual vector valued duality we construct vector valued distributions
(linear continuous functionals) which will be denoted again by $\mathcal{X} $. The last construction has nothing
to do with Grassmann integrals but $\mathcal{X}$ constructed by the above mentioned supersymmetric duality and
$\mathcal{X} $ constructed with the help of the vector valued duality are certainly related. The relation is

\begin{gather}
\mathcal{X}_V (X_{inv}) =\mathcal{X}_G (X)
\end{gather}
where the index $V$ stays for vector valued, $G$ for additional Grassmann integration in the duality functional
and the subscript inv indicates an involution of the coefficients of $X$. It exchanges $f$ with $d$, then $\varphi
,\chi $ with $\lambda ,\psi $ and finally keeps $v$ unchanged. It is a direct consequence of the Grassmann
integration in the r.h.s. of (2.13) which selects coefficients of the highest power $\theta^2 \bar \theta^2 $. We
do not further insist on this relation because as we will see it is only of limited help for us. \\ Concluding
there are at least three possibilities to look at superdistributions. We argued that these possibilities are in
principle related  but we didn't went into further technical details. Our discussion is nevertheless sufficient in
order to point out some difficulties which appear in the general framework of superdistributios. These
difficulties are related to complex (and Grassmann) conjugation and to tensor products. Defining complex
conjugation of usual distributions and related them to a Hilbert space structure (as this is necessary for
instance in the discussion of the Gelfand triple) poses no problems. On the contrary complex (and Grassmann)
conjugation of superdistributions in the duality interpretation may pose problems because Grassmann conjugation
reverses the order of the Grassmann variables. We will avoid such problems from the beginning considering only
superdistributions which are even as polynomials in the Grassmann variables.

\section{Nuclearity in supersymmetric quantum field theory}

Our interest is devoted now to the second difficulty, i.e. to supersymmetric tensor products. Consider the product
$\mathcal{X}_1 (X_1)\mathcal{X}_2 (X_2)$ where $X_1 ,X_2 $ are superfunctions in different variables and
$\mathcal{X}_1 ,\mathcal{X}_2 $ superdistributions of type (2.1). Then for $ \mathcal{X}=\mathcal{X}_1
\mathcal{X}_2 $ we generally have

\begin{gather}
\mathcal{X}(X_1 X_2 )\neq \mathcal{X}_1 (X_1 )\mathcal{X}_1 (X_2 )
\end{gather}
This happens because $X_1 $ generally does not commute with $\mathcal{X}_2 $ (remember that our "spinorial"
coefficients being functions or distributions commute with Grassmann variables). In this case there may be some
problems with the distribution theoretic tensor products. Such problems do not appear if for instance $
\mathcal{X}_2 $ contains only even powers of the Grassmann variables. In this case

\begin{gather}
\mathcal{X}(X_1 X_2 )=\mathcal{X}_1 (X_1 )\mathcal{X}_2 (X_2 )
\end{gather}
Topological aspects of tensor products can be inferred from the vector valued interpretation or by introducing
from the beginning as mentioned above seminorms (norms) directly on functions of type (2.1) with indefinitely
differentiable coefficients. Nuclearity (i.e. the kernel theorem well known in distribution theory) seems to be
assured in the vector valued interpretation but it is not what we need. Let us discuss this property in the
physical framework of supersymmetric quantum field theory. Nuclearity is needed in quantum field theory mainly by
introducing the n-point functions as distributions in their joint variables. Indeed if $V$ is the quantum field
operator considered as operator valued distribution \cite{SW}, $\Omega $ the vacuum and $X_1 ,X_2 \ldots ,X_n $
test functions in $S(R^4 )$ then $(\Omega ,V(X_1 )V(X_2 )...V(X_n )\Omega $) is linear and separately continuous
in $X_1 ,X_2 \ldots ,X_n $. By nuclearity it defines the distribution $W_n (x_1 ,x_2 ,\ldots ,x_n )\in S'(R^{4n})$
in 4n-variables $x_1 ,x_2 ,\ldots ,x_n $ such that $(\Omega ,V(X_1 )V(X_2 )...V(X_n )\Omega )=W_n (X_1 ,X_2
,\ldots ,X_n )$. In the supersymetric quantum field theory some problems of similar nature as in the previous
section may appear because of noncommutativity of $V(z_i)$ and $X_j (z_j )$. In order to avoid them we may assume
at a certain stage that $(\Omega ,V(X_1 )V(X_2 )...V(X_n )\Omega $) can be written as multilinear form $W_n (X_1
,X_2 ,\ldots ,X_n )$ separately continuous in $X_1 ,X_2 \ldots ,X_n $. This is the case if  the $n$-point
functions are even in the Grassmann variables. Then by nuclearity $W_n =W_n(z_1 ,z_2 ,\ldots ,z_n )\in
S'(R^{4n}\times G )$. Certainly we can avoid discussing nuclearity at all by assuming that the $n$ point functions
$W_n =W_n(z_1 ,z_2 ,\ldots ,z_n )\in S'(R^{4n}\times G )$ with
\begin{gather}
(\Omega ,V(X_1 )V(X_2 )...V(X_n )\Omega )=W_n (X_1 ,X_2 ,\ldots ,X_n )
\end{gather}
exist as superdistributions. This assumption is very natural because it reflects and captures formal computations
in physics. For example computing the two-point function of the free Wess-Zumino model (directly or by means of
functional integrals) we use fields with noncommuting spinor coefficients. This is equivalent in this particular
case with one form or another of the above mentioned assumption. Moreover the two-point functions turn out to be
even in Grassmann variables. The $n$-point functions of the free field are products of (even) two point functions
and pose no problems either. Generally in the next sections we will prove that supersymmetric invariance in the
scalar case implies that the $n$-point functions are even in the Grassmann variables. More details including
further examples and their relation with the field reconstruction theorem will appear in the next sections.
\\ At this point, as a further technical remark, it is worthwhile to anticipate some points of our study remembering
first that elementary nuclearity is related to the idea of the Gelfand rigged space (Gelfand triple). The most
well known Gelfand triple is $ S \subset L^2 \subset S^{'} $ where $S^{'} $ is the space of tempered distributions
over the test function space $S$ of rapidely decreasing functions. In supersymmetry we generally do not have this
situation because (as we will see) the SUSY $L^2 $ space off-shell doesn't exist. On the other hand we will show
that the supersymmetric (invariant) $L^2 $ space do exists on shell being segregated by a Krein structure of the
off-shell superspace. A key point will be the fact that general distribution theoretic tools off shell, as
discussed above, will be compatible with the Hilbert space (including Gelfand triple) considerations on shell.
\\
We conclude that possible difficulties with superdistributions related to complex conjugation and especially to
tensor products and nuclearity do not show up in supersymmetric quantum field theory. This encourages us to look
for other structures in the following sections.

\section{Supersymmetric *-algebra}

For the time being we stay off shell and introduce the counterpart of a topological *-algebra used in the
reconstruction theorem \cite{SW}. This is a locally convex *-algebra constructed with the help of supersymmetric
functions. Suppose that the coefficients of the supersymmetric functions $X(z_1 ,z_2 ,\ldots ,z_n )$ belong to the
Schwartz space $S(R^{4n})$. Let $\mathcal{B}(S)$ be the set of sequences $(X^n ),n=0,1,2,\ldots $ such that $X^0 $
is a complex number and $X^n =0$ for all but finite many n's. In the vector space $\mathcal{B}(S)$ we define the
operations of multiplication and the *-operation:

\begin{gather}
(XY)^n (z_1 ,z_2 ,\ldots ,z_n )=\sum_{j=0}^n X^j (z_1 ,\ldots ,z_j )Y^{n-j}(z_{j+1},\ldots ,z_n ) \\ (X^* )^n (z_1
,\ldots ,z_n )=\overline {X^n (z_n ,\ldots ,z_1 )}
\end{gather}
where bar means at the same time complex and Grassmann conjugation. We prove that $*$ is an involution. This
operation make $\mathcal{B}(S)$ into an algebra with involution and identity. It is not a Banach algebra but it
can be turn as usual into a locally convex algebra. Positivity and positive definite functionals (states) on this
algebra will play a central role. At this point let us remark that a priori it is not easy to find states on the
supersymmetric algebra $\mathcal{B}(S)$ because the presence of Grassmann variables. The problem will be studied
later in the paper. We come back to the involution properties. It is clear that $*$ is conjugate antilinear. It
remains to prove that it reverses the order of products $(XY)^* =Y^* X^* $ and that $(X^*)^* =X $. This follows
from the relation

\begin{gather}
\overline{X^k (z_1 ,\ldots ,z_k )Y^l (w_1 ,\ldots ,w_l )}=\overline{Y^l (w_1 ,\ldots ,w_l )}\quad \overline{X^k
(z_1 ,\ldots ,z_1 )}
\end{gather}
This algebra will be the framework of the reconstruction theorem which provide us with supersymmetric
operator-valued fields. Note that for usual functions i.e. functions which do not depend on Grassmann variables we
have besides (4.3) also $\overline{X^k Y^l}=\bar X^k \bar Y^l  $, a relation which is generally invalid if
Grassmann variables are present and the bar includes Grassmann conjugation. But this relation is not used in the
Gelfand-Neumann like construction of the reconstruction theorem which we will perform in this paper. The crucial
hermiticity of the scalar product to be constructed rests only on the relation $(XY)^* =Y^* X^* $ which holds in
both usual and supersymmetric framework. This shows that supersymmetry is compatible with a quantum field theory
$C^* $-approach and makes use of the *-algebra structure in an even more intricate way. Indeed the involution * in
supersymmetry includes Grassmann conjugation besides the complex one. \\ The reconstruction theorem itself in the
supersymmetric framework does not seem to raise any special problems as this was already noted in \cite{O}. But at
this stage the reader may feel uneasy: the positivity needed in the reconstruction theorem can be easily
formulated using the *-algebra above but it is not clear to what extent such a property (i.e. positivity) can be
realized in a framework in which Grassmann variables and Berezin integration play a central role. Experience with
Berezin integration shows that it is not easy to find sound positive definite sesquilinear forms satisfying
hermiticity (scalar products) given directly on functions depending of Grassmann variables, i.e. analog to $L^2 $
scalar products, although such examples exist \cite{RY}. At this point invariance under the supersymmetric
Poincare group helps and this matter is the subject of the next sections. In \cite{O} a detour over the group
(Hopf) algebra was proposed which we do not follow here.

\section{Hilbert Spaces of Supersymmetric Functions}

In this section we present off-shell pre-Hilbert spaces and on-shell Hilbert spaces of functions in supersymmetry
which are related to the $L^2 $ and Sobolev spaces of analysis. It means that we look for positive sesquilinear
forms by integrating superfunctions in all variables including the Grassmann ones. It is clear that such
sesquilinear forms in view of Berezin integrations must be nontrivial. Let $X,Y$ be supersymmetric functions of
type (2.1) with regular coefficients. The key sesquilinear form is

\begin{gather}\nonumber
(X,Y)=\int d^8z_1d^8z_2\bar X(z_1)K(z_1 ,z_2)Y(z_2)= \\ \nonumber =\int d^8z_1d^8z_2\bar X(z_1)[(P_c +P_a -P_T
)K_0 (z_1-z_2)]Y(z_2)= \\ =\int d^8z_1d^8z_2\bar X(z_1)(P_c +P_a -P_T )K_0 (z_1-z_2) Y(z_2)
\end{gather}
where $P_i ,i=c,a,T $ are the chiral, antichiral and trasversal projections respectively \cite{WB,S} acting on the
$z_1 $-variable (see equations to follow) and

\begin{gather} \nonumber
K_0 (z)=\delta^2 (\theta )\delta^2 (\bar \theta )D^+ (x) \\ D^+ (x)=\int e^{ipx}d\rho (p) =\\ =\int e^{-ipx}d\rho
(-p)
\end{gather}
We have used the standard notations

\begin{gather}
D_{\alpha }=\partial_{\alpha } +i\sigma_{\alpha \dot \alpha }^l\bar \theta^{\dot \alpha }\partial _l \\ D^{\alpha
}=\epsilon ^{\alpha \beta }D_{\beta }=-\partial ^{\alpha }+i\sigma^{l\alpha }_{\dot \alpha }\bar \theta^{\dot
\alpha }\partial _l  \\ \bar D_{\dot \alpha }=-\bar {\partial }_{\dot \alpha } -i\theta ^{\alpha }\sigma _{\alpha
\dot \alpha }^l\partial _l \\ \bar D^{\dot \alpha }=\epsilon ^{\dot \alpha \dot \beta }\bar D_{\dot
  \beta }=\bar {\partial }^{\dot \alpha }-i\theta ^{\alpha }\sigma
_{\alpha }^{l\dot \alpha }\partial _l
\end{gather}
for the covariant (and invariant) derivatives and

\begin{gather}
D^2=D^{\alpha }D_{\alpha }=-(\partial ^{\alpha }\partial _{\alpha }-2i\partial _{\alpha \dot \alpha }\bar \theta
^{\dot \alpha } \partial ^{\alpha }  +\bar \theta^2 \square ) \\ \bar D^2=\bar D_{\dot \alpha }\bar D^{\dot \alpha
}=-(\bar {\partial } _{\dot \alpha }\bar {\partial }^{\dot \alpha }+2i \theta ^{\alpha }\partial _{\alpha \dot
\alpha }\bar \partial ^{ \dot \alpha } +\theta^2\ \square )
\end{gather}

\begin{gather}
c=\bar D^2D^2, a=D^2\bar D^2, T=D^{\alpha }\bar D^2 D_{\alpha }=\bar D_{\dot \alpha }D^2 \bar D^{\dot \alpha
}=-8\square +\frac {1}{2}(c+a)
\end{gather}

\begin{gather}
P_c=\frac {1}{16\square }c,P_a=\frac {1}{16\square }a, P_T=-\frac {1}{8\square }T
\end{gather}
The positive measure $d\rho (p) $ is concentrated in the interior of the forward light cone ( $d\rho (-p)$ is
concentrated in the interior of the backward light cone). It can be proved by direct computation \cite {C1} (see
also the remark viii) later in this section and the section 7 of this paper) that (5.1) is positive definite. It
is not strictly positive definite because it may have plenty of zero vectors. The well-known example is $d\rho (p)
=\theta (p_0 )\delta (p^2 +m^2 ), d\rho (-p)=\theta (-p_0 )\delta (p^2 +m^2 ) $ where $m$ is the mass (the
"massive" case). In the massive case the zero vectors are harmless because they can be easily eliminated by the
on-shell condition; the "massless" case require a certain amount of extra work \cite{C1} but it is not too much
different from the massive one. We can also write

\begin{gather}\nonumber
(X,Y)=\int d^8z_1d^8z_2 [(P_c +P_a -P_T )K_0 (z_1-z_2)] \bar X(z_1 )Y(z_2)= \\ =\int d^8z_1d^8z_2 (P_c +P_a -P_T
)K_0 (z_1-z_2) \bar X(z_1 )Y(z_2)
\end{gather}
because $(P_c +P_a -P_T )K_0 $ is even in the Grassman variables. \\ Beside being positive (for $X=Y$) the form
(.,.) satisfy the usual (hermiticity) property

\begin{gather}
(X,Y)=\overline{(Y,X)}
\end{gather}
where the bar on the r.h.s. means complex conjugation. Moreover we have

\begin{gather}
\overline{(Y,X)}=(\bar Y,\bar X)
\end{gather}
where the bars on the r.h.s. in (5.14) mean complex conjugation supplemented by Grassmann conjugation. The proofs
including the positivity (i.e. the non-negativity of (.,.)) are by computation.\\ Some remarks are in order : \\

i) $X,Y$ are functions (not fields). Their spinorial components are assumed to commute between themselves and with
the Grassmann variables. Consequently we apply the mixed van de Waerden caculus which was worked out in \cite{C1}
instead of the usual one.
\\

ii) There is a surprising minus sign in front of $P_T $ in (5.1). Reversing it produces an invariant, indefinite
(not only semidefinite) sesquilinear form

\begin{gather}
<X,Y>=\int d^8z_1d^8z_2\bar X(z_1)K_0 (z_1-z_2) Y(z_2)
\end{gather}
because $P_c +P_a +P_T =1 $. This is the indication of an intrinsec  Krein structure of the $N=1$ superspace. The
associated Hilbert space will be introduced bellow. \\

iii) The operators $P_c ,P_a ,P_T $ in (5.1) can be moved to act on the $z_2 $-variable.
\\

iv) the integral kernel in (5.15) depends only on $z_1 -z_2 $ whereas the full integral kernel in (5.1) does not
share this property. \\

v) Our positive sesquilinear form (5.1) is different from other proposals in supersymmetry (see \cite{DW} as well
as \cite{F} and the references given there) known to the author which seem to have the peculiar property of
generating complex-valued norms.
\\

vi) It is interesting to remark that the results of the relevant computation (like for instance the proof of
positivity of (5.1) and hermiticity (5.13)) is the same for $X,Y$ being functions or fields. Whereas the results
for functions are rigorous, the ones for fields, although computationally correct, must be looked at as formal. \\

vii) The operation of complex and Grassmann conjugation taken together is a conjugation operator in the Hilbert
space.
\\

viii) An equivalent expression for the scalar product $(.,.)$ is

\begin{gather} \nonumber
(X_1 ,X_2 )=\int d^4 z_1 d^4 z_2 [(P_c +P_a +P_T )\bar X(z_1 )]X(z_2 )= \\ \nonumber =\int d^4 x_1 d^4 x_2 D^+(x_1
-x_2 )\times
\\  \nonumber \times \int d^2 \theta d^2 \bar \theta \bar X_1 (x_1 ,\theta, \bar \theta)[(P_c +P_a -P_T )X_2 ](x_2
,\theta, \bar \theta )= \\  =\int d^4 x_1 d^4 x_2 D^+(x_1 -x_2 )(I_c +I_a -I_T )
\end{gather}
where the notations are self-explanatory. This form is convenient for computation, in particular for those needed
in section 7. \\

ix) Not only the supersymmetric invariant kernel induced by $P_c +P_a -P_T $ produces a positive result but also
separately $P_c ,P_a $ and $ -P_T $. Moreover positivity is also induced by $P_c +P_a +P_+ +P_- $ in the notation
of \cite{WB}. The last property is connected to the positivity of the two-point function for the free Wess-Zumino
model \cite{C1} in which the trasversal sector is neglected.  \\

Finally we factorize the zero-vectors of the positive sesquilinear form (5.1) and by completion obtain our Hilbert
space. This factorization is for the case  $d\rho (p) =\theta (p_0 )\delta (p^2 +m^2 )dp, m>0 $ equivalent with
the on-shell restriction $p^2 =m^2 $. Altogether we generated an inherent Hilbert-Krein structure on $N=1$
supersymmetric functions which we call the standard Hilbert-Krein structure of the superspace in order to
distinguish it from other ones. The deceptive simple point of the whole business (which intuitively supports this
paper) seems to be the minus sine in front of the transversal projection! The inherited bona fide Hilbert space is
the analog of the (on-shell) invariant $L^2 $-space. It is not a non-Hilbert generalization of a Hilbert space
sometimes used in supersymmetries in which vectors can have complex lengths \cite{F}. It is likely that it can be
used in applications as for instance the Wigner type representation theory on supersymmetric functions,
supersymmetric canonical quantization (for first steps see \cite{C2}) and renormalization including gauge
theories.

\section{Supersymmetric Hilbert space operators and their adjoints}

Now we come to supersymmetric operators and especially to theis adjoints. Usual examples of formal supersymmetric
operators are multiplication and especially differential (covariant and invariant) operators of the form
\[ D_{\beta },\bar D_{\dot \beta }, D^2 ,\bar D^2 ,c=\bar D^2 D^2, a=D^2 \bar D^2 ,T=D^{\beta }\bar D^2 D_{\beta
}=\bar D_{\dot \beta }D^2 \bar D^{\dot \beta } \]  etc. as well as analog operators (supersymmetric generators)
$Q_{\beta },\bar Q_{\dot \beta } $ etc. Before starting let us compute

\begin{gather} \nonumber
D_\beta X=\varphi_\beta +\theta^\alpha (2m\epsilon_{\beta \alpha })+\bar \theta_{\dot \alpha }(-v_\beta ^{\dot
\alpha }-i\sigma_\beta^{l\dot \alpha}\partial_l f)+ \\ \nonumber +\bar \theta^2 (\psi_{\beta }
-\frac{i}{2}\sigma_{\beta \dot \beta }^l \partial_l \bar \chi^{\dot \beta })+\theta^\alpha \bar \theta^{\dot
\alpha }(2\epsilon_{\alpha \beta }\bar \lambda_{\dot \alpha }-i\sigma_{\beta \dot \alpha }^l \partial_l
\varphi_\alpha )+  \\ +\theta^2 \bar \theta_{\dot \alpha }(-i\sigma_\beta^{l\dot \alpha }\partial_l m)+\bar
\theta^2 \theta^\alpha (2\epsilon_{\beta \alpha }d+\frac{i}{2}\sigma_\beta^{l\dot \beta }\partial_l v_{\alpha
\dot\beta })-\frac{i}{2}\theta^2 \bar \theta^2 \sigma_{\beta \dot \beta }^l \partial_l \bar \lambda^{\dot \beta }
\end{gather}
and a similar expression for $\bar D_{\dot \beta }X $. \\ Note first that $D_\beta ,\bar D_{\dot \beta } $ are
true operators applied to functions $X$ but no definite mathematical objects when applied to fields because for
instance $2m\epsilon_{\alpha \beta }$ are no anticommuting spinor components (for fixed $\beta $). A similar
remark applies to the $Q$-operators. This doesn't affect us because we work in the frame of test functions in
which the "fermionic" components commute. \\ We come to supersymmetric adjoint operators. Generally the Hilbert
space adjoint $A^+ $ of an operator $A$ is (here and further on in this paper we disregard domain problems because
as usual in elementary quantum field theory they turn out to be harmless) defined through

\begin{gather}
(X,AY)=(A^+ X,Y)
\end{gather}
This will be our definition too. It uses the supersymmetric true scalar product $(.,.)$. As in \cite{GGRS}, using
the real bilinear form associated to (5.1), the transpose and the conjugate of an operator can be introduced but
we do not need them. \\ We pass now to the most common examples. In order to study them we need some partial
integration results \cite{S}. We have

\begin{gather}
\int d^8 z\bar XD_\beta Y=\mp \int d^8 z(D_\beta \bar X)Y \\ \int d^8 z\bar X\bar D_{\dot \beta }Y=\mp \int d^8
z(\bar D_{\dot \beta }\bar X)Y
\end{gather}
according as $\bar X$ (or $X$) is an even or odd Grassmann function. On the other hand we have from our previous
considerations for functions

\begin{gather}
\overline {D_\beta X}=\mp \bar D_{\dot \beta }\bar X
\end{gather}
according as $\bar X$ (or $X$) is even or odd. Consequently

\begin{gather}
\int d^8 z\bar XD_\beta Y=\int d^8 z(\overline{\bar D_{\dot \beta }X})Y \\ \int d^8 z\bar X\bar D_{\dot \beta
}Y=\int d^8 z(\overline{D_\beta X})Y
\end{gather}
for arbitrary $X,Y$. Similar relations (6.3)-(6.7) hold for the operators $Q_{\beta }, \bar Q_{\dot \beta } $ to
be defined later in this section. Now we insert the kernels in the integrals. Indicating by superscrips the acted
variables we have

\begin{gather}\nonumber
D^{2(1)}(P_c +P_a -P_T )^{(1)}K_0 (z_1 -z_2 )= \\ =D^{2(2)}(P_c +P_a -P_T )^{(1)}K_0 (z_1 -z_2 )\\  \nonumber \bar
D^{2(1)}(P_c +P_a -P_T )^{(1)}K_0 (z_1 -z_2 )= \\ =\bar D^{2(2)}(P_c +P_a -P_T )^{(1)}K_0 (z_1 -z_2 )
\end{gather}
Using (6.3)-(6.9) we get

\begin{gather}
(X,D^2 Y)=(\bar D^2 X,Y) \\ (X,\bar D^2 Y)=(D^2 X,Y)
\end{gather}
i.e. $\bar D^2 $ is the operator adjoint of $D^2 $ and vice versa:

\begin{gather}
(D^2)^+ =\bar D^2,\quad (\bar D^2)^+ =D^2
\end{gather}
On the same lines we obtain as bona fide Hilbert space operators

\begin{gather}
c^+ =c, a^+ =a, T^+ =T
\\   P_c ^+ =P_c ,P_a^+ =P_a ,P_T^+ =P_T
\end{gather}
as well as adjoint relations for the supersymmetric operator generators $Q_{\beta}^+ =\bar Q_{\dot \beta } ,(\bar
Q_{\dot \beta})^+ =Q_\beta $ etc. The operators $ Q_{\alpha }, \bar Q_{\dot \alpha } $ are given by

\begin{gather}
Q_{\alpha }=\partial_{\alpha } +i\sigma_{\alpha \dot \alpha }^l\bar \theta^{\dot \alpha }\partial _l \\ \bar
Q_{\dot \alpha }=-\bar {\partial }_{\dot \alpha } -i\theta ^{\alpha }\sigma _{\alpha \dot \alpha }^l\partial _l
\end{gather}
Explicitly we have

\begin{gather}\nonumber
Q_\beta X=\varphi_\beta +\theta^\alpha (2m\epsilon_{\beta \alpha })+\bar \theta_{\dot \alpha }(-v_\beta ^{\dot
\alpha }+i\sigma_\beta^{l\dot \alpha}\partial_l f)+ \\ \nonumber +\bar \theta^2 (\psi_{\beta }
+\frac{i}{2}\sigma_{\beta \dot \beta }^l \partial_l \bar \chi^{\dot \beta })+\theta^\alpha \bar \theta^{\dot
\alpha }(2\epsilon_{\alpha \beta }\bar \lambda_{\dot \alpha }+i\sigma_{\beta \dot \alpha }^l \partial_l
\varphi_\alpha )+
\\ +\theta^2 \bar \theta_{\dot \alpha }(i\sigma_\beta^{l\dot \alpha }\partial_l m)+\bar \theta^2 \theta^\alpha
(2\epsilon_{\beta \alpha }d-\frac{i}{2}\sigma_\beta^{l\dot \beta }\partial_l v_{\alpha \dot\beta
})+\frac{i}{2}\theta^2 \bar \theta^2 \sigma_{\beta \dot \beta }^l \partial_l \bar \lambda^{\dot \beta }
\end{gather}
and a similar relation for $\bar Q_{\dot \beta}$. \\ Introducing the $\theta ,\bar \theta $-expansion for $X$ we
get the action of $Q,\bar Q $ on components which is well-known in elementary supersymmetry. \\ In order to prove
the operator adjoint relations $Q_{\beta }^+ =\bar Q_{\dot \beta } ,(\bar Q_{\dot \beta })^+ =Q_{\beta } $ we use
the commutativity of $Q$ and $D$ operators and

\begin{gather}\nonumber
Q_\beta ^{(1)}(P_c +P_a -P_T )^{(1)}K_0 (z_1 -z_2 )= \\ =-Q_\beta^{(2)}(P_c +P_a -P_T )^{(1)}K_0 (z_1 -z_2 )\\
\nonumber \bar Q_{\dot \beta }^{(1)}(P_c +P_a -P_T )^{(1)}K_0 (z_1 -z_2 )= \\ =-\bar Q_{\dot \beta }^{(2)}(P_c
+P_a -P_T )^{(1)}K_0 (z_1 -z_2 )
\end{gather}
Relations of type (6.18),(6.19) do not hold for the $D_{\beta },\bar D_{\dot \beta } $ operators . This is the
reason we cannot simply relate the $D_{\beta }$ to the $\bar D_{\dot \beta } $ operators in our Hilbert space
understanding.
\\
If $P$ is the momentum operator than $P, \xi Q+ \bar \xi \bar Q ,aP+\xi Q+\bar \xi \bar Q $ are self adjoint
operators. Consequently the operator $\exp {i(aP+\xi Q+\bar \xi \bar Q}$ is an unitary operator ($ U^+ =U^{-1},
(UX,UY)=(X,Y) $). Here we use the fact that $Q$ commutes with $P$ and $D, \bar D $. If $\tau =\tau (a,\xi ,\bar
\xi )$  is the supersymmetric transformation in superspace than we can implement it unitarily on supersymmetric
functions by the usual formula

\begin{gather}
U(\tau )X(z)=X(\tau(z))
\end{gather}
This is the rigorous version of a well known formal statement (\cite{W},p.91; see also \cite{BK}). If a Lorentz
spin is present this formula changes as in \cite{W}. Generally it is possible to construct the (irreducible)
unitary representations of supersymmetry on functions (not on fields in which case the unitarity is formal
\cite{W,BK}) as this originally appears in the Wigner-Makey theory for the Poincare group. Measure-theoretic
considerations in the Makey theory of induced representations should be replaced by analog considerations in the
framework of von Neumann algebras. \\ Before ending this section let us point out that the supersymmetric
framework can provide sometimes surprising results. Indeed being unitary the operator $ exp{i(\xi Q+\bar \xi \bar
Q )} $ studied above is bounded. But expanding it in powers of $\xi ,\bar \xi $ it produces a sum of a finite
number of powers of $Q ,\bar Q $ which at the first glance seems to be unbounded. Though there is no contradiction
because the boundnes is a consequence of the supersymmetric framework. This was already noted in \cite{O}.

\section{Hilbert space realization on (multiplet) components}

The supersymmetric fields appear in physics as multiplets of ordinary quantum fields, bosons and fermions. Whereas
the supersymmetry imposes some conditions on the ordinary fields in the multiplet (for instance same number of
bosons as fermions), the idea (at least in the perturbative framework) is to start with free multiplets on which
an interaction (in form of a interaction Langragian) is superimposed. The treatement makes use of supersymmetric
technique or equivalently, if such a technique doesn't exist or is insuficiently developed, the work goes directly
on component fields. At the first sight the Hilbert space (Fock space) of the free theory appears as a tensor
product of the multiplet components. But this picture is too simple in order to be true. Indeed supersymmetry
induces relations between multiplet components and a simple tensor product picture has to be aborted. On the other
hand being in the free case we expect nevertheless a kind of Fock-type space. How does it looks like? This is the
question which we do answer in this section even in a more general setting. We start with the standard Hilbert
space of the $N=1$ supersymmetry introduced in section 5 and decompose it on components. This is a matter of
computation and we will give here only the results and some hints. The easiest way is to use (5.16) where we
compute

\begin{gather} \nonumber
I_c +I_a -I_T = \\ \nonumber =\frac{1}{2}\bar \varphi_1 i\bar \sigma^l \partial_l \varphi_2 +\frac{1}{2}\chi_1
i\sigma^l
\partial_l \bar \chi_2 +\frac{2}{\square}\lambda_1 i\sigma^l \partial_l \bar \lambda_2 +
\frac{2}{\square}\psi_1 i\bar \sigma^l \partial_l \psi_2 +\\ +\frac{\square}{4}\bar f_1 f_2 +\bar m_1 m_2 +\bar
n_1 n_2 +\frac{4}{\square}\bar d_1 d_2 -\bar v_1^l \partial_l
\partial_k v_2^k +\frac{\square }{2}\bar v_{1l}v^{2l}
\end{gather}
where  $f_1 (x_1 )=f_1, f_2 (x_2 )=f_2 ,\varphi_1 (x_1 )=\varphi_1 ,\varphi_2 (x_2 )=\varphi_2 $ etc. We take $X_1
=X_2 $ which means $f_1 =f_2 ,\varphi_1 =\varphi_2 $ etc. The contributions involving $f,d,m,n$ in (7.1) are
obviously positive (i.e. non-negative). The same holds for the $v$-contribution after partial integration. We have
only to look at the $\varphi ,\chi ,\lambda $ and $\psi $-contrinutions. Let

\begin{gather}
J_1 =\int d^4 x_1 d^4x_2 D^+ (x_1 -x_2 )\bar \varphi_1 (i\bar \sigma^l \partial_l )\varphi_2  \\ J_2 =\int d^4 x_1
d^4x_2 D^+ (x_1 -x_2 )\chi_1 (i\sigma^l \partial_l )\bar \chi_2
\end{gather}
In order to study $J_1 ,J_2 $ we go to the Fourier transform defined as
\[ \tilde f(p)=\frac{1}{(2\pi )^2 }\int e^{-ipx}f(x)d^4 x \]
obtaining
\begin{gather}\nonumber
J_1 =\int d^4 p \tilde {\bar \varphi }_1 (p)\tilde D^+ (p)i\bar \sigma \tilde {\partial_l \varphi_2 }(-p)= \\
=\int d^4 p\overline {\tilde \varphi_1 }(-p)\rho (-p)i\bar \sigma^l (ip_l)\tilde {\varphi_2 }(-p)=\int d^4
p\overline {\tilde \varphi_1 }(p)\rho (p)(\bar \sigma^l p_l)\tilde {\varphi_2 }(p)
\end{gather}
where a Fourier normalization factor was omited and we have used $\tilde {\bar f}(-p)=\bar {\tilde f}(p)$ and
$\int d^4 x_1 d^4 x_2 D^+ (x_1 -x_2 )F(x_1 )G(x_2 )=\int d^4 p \tilde F (p) \tilde D^+ (p)\tilde G(-p) $ where
$F,G$ are arbitrary functions. But $\bar \sigma^l p_l $ is positive in the forward light cone where $d\rho (p)$ is
concentrated such that (for $\varphi_1 =\varphi_2 )$ $J_1 $ is positive. In the same way $J_2 $ is positive
because $\sigma^l p_l $ is positive in the forward light cone too. Similar arguments hold for the term in $\lambda
,\psi $ in (7.1). We conclude that (5.16) is  for $X_1 =X_2 $ positive (non-negative). The hermiticity relations
(5.13),(5.14) follow taking into account that $\sigma ,\bar \sigma $ are selfadjoint. \\ Adopting appropiate
normalizations we obtain the direct sum

\begin{gather}
H=\oplus H_{components}
\end{gather}
where $H$ is the Hilbert space of the $N=1$ superspace and $H_{components}$ the Hilbert spaces of the components
(of bosonic and fermionic nature). The fermionic components (Weyl or Majorana) do not involve the mass term of
their corresponding two point function. The formula (7.5) simplifies for chiral, antichiral and transversal
sectors. This seems to be an interesting result because it says that the standard supersymmetric Hilbert space is
(because of supersymmetric invariance) a direct sum (not a direct product!) over the components. For a specific
model like the vector field or the free Wess-Zumino model the elements in $H$ are the "one particle" states and
the Fock space is constructed as usual. It is not clear to us to what extent this realization of the Fock space is
or not of some use and this is the reason we do not continue on this line. Acting in $H$ the supersymmetric
operators, as expected, mix up the components. An example are the generators $Q,\bar Q $. Working out their action
in (7.5) we re-obtain the usual supersymmetric transformations on multiplet components.

\section{The simplest
example: free supersymmetric quantum field}

The first model of a supersymmetric quantum field theory in four dimensions was the chiral-antichiral Wess-Zumino
model with a third power interaction. Without the interaction the Wess-Zumino (free) field is still a
chiral-antichiral one. We prefer instead of the Wess-Zumino model the free massive vector field $V(z)$ which we
define here by the simplest supersymmetric invariant two point function which at the same time is positive
definite. The corresponding kernel in the sense of (5.1) is induced by $P_c +P_a -P_T $ (the generalization
$\zeta_c P_c +\zeta_a P_a -\zeta_T P_T $ with $\zeta_i \geq 0$ is also posible). We start by defining
representations of supersymmetries on supersymmetric functions by

\begin{gather}
U(\tau)X(z)=X(\tau (z))
\end{gather}
where $\tau =\tau (z)$ is a supersymmetric transformation. Considered in the Krein space, i.e. in the space with
inner product $<.,.> $, as this is usually suggested in the literature on the subject \cite{W,BK}, this
representation is not unitary. We obtain a unitary representation when representing in the corresponding Hilbert
space $ H $ with scalar product $(.,.)$. The supersymmetric quantum field $V(z)$ is similar to the scalar neutral
field in the common quantum field theory. Its definition requires a (supersymmetric) symmetric Fock  space. Indeed
we construct the symmetric Fock space on $H $ which we denote by $\mathcal{F}=Fock(H)$ (antisymmetric Fock spaces
are reserved for ghosts). A general element of $\mathcal{F} $ will be denoted by $\Phi =(\Phi^{(0)}
,\Phi^{(1)},\ldots \Phi^{(n)}\ldots ), \Phi^{(0)}=1, \Phi^{(n)}=\Phi^{(n)}(z_1 ,z_2 \ldots ,z_n ) $. Note that
working with the overall anticommuting convention discussed above, Fock spaces in supersymmetry in the scalar case
(complying with the right statistics) are always symmetric. The above unitary representation can be extended as
usually to the Fock space $\mathcal{F}$. We set for the vector field $V$ smeared with the (test) supersymmetric
function $X(z)$:

\begin{gather}
V(X)=V^+(X)+V^-(X)
\end{gather}
with

\begin{gather}
(V^+(X)\Phi )^{(n)}(w_1 ,\ldots ,w_n )=\sqrt{n+1}(X(w),\Phi^{(n+1)}(w,w_1 ,\ldots ,w_n )) \\ (V^-(X)\Phi
)^{(n)}(w_1 ,\ldots ,w_n )=\frac{1}{\sqrt n}\sum_{j=1}^n X(w_j )\Phi^{(n-1)}(w_1,\ldots ,\hat w_j \ldots ,w_n )
\end{gather}
where $w=(p,\theta ,\bar \theta )$ and $p$ is the momentum. The adjoint of $V$ is given by $V^+ (X)=V(\bar X )$,
i.e. we have a real (neutral) supersymmetric field. We can project on components using the representation (7.5) of
the Hilbert space $ H $ but we are not interested in this question. The n-point functions can be given as usual as
products of two-point functions. They are non vanishing only for even $n$ and are even functions of the Grassmann
variables in the combination $\theta \bar \theta $. This follows from the relation (see for instance \cite{C2})

\begin{gather} \nonumber
(P_c +P_a -P_T )\delta^2 (\theta_1 -\theta_2 )\delta^2 (\bar \theta_1 -\bar \theta_2 )= \\ \nonumber =(1-2P_T
)\delta^2 (\theta_1 -\theta_2 )\delta^2 (\bar \theta_1 -\bar \theta_2 )=\\ \nonumber =\frac{4}{\square
}(1-i\theta_1 \sigma^l \bar \theta_2 \partial_l -i\bar \theta_1 \bar \sigma^l \theta_2 \partial_l )+\theta_1 ^2
\bar \theta_2 ^2 +\bar \theta_1 ^2 \theta_2 ^2 + \\ \nonumber  +2(\theta _1 \sigma^l \bar \theta_1 )(\theta _2
\sigma_l \bar \theta_2 +\frac{2}{\square }\partial_l \partial^m \bar \theta_2 \bar \sigma_m \theta_2 )- \\
-i\theta_1 ^2 \bar \theta_1 \bar \sigma^l\theta_2 \bar \theta_2 ^2 \partial_l -i\bar \theta_1 ^2 \theta_1 \sigma^l
\bar \theta_2 \theta_2 ^2
\partial_l +\frac{1}{4}\square \theta_1 ^2 \bar \theta_1 ^2 \theta_2 ^2 \bar \theta_2 ^2
\end{gather}
This last property which in fact, because of supersymmetric invariance, holds in any supersymmetric quantum field
theory (see the next section) makes obsolete the problems connected to nuclearity which have been discussed in
section .

\section{Implications of supersymmetric invariance to
the $n$-point functions}

In this section we give the general form of a supersymmetric invariant $n$-point function generalizing a result in
\cite{C2}. Suppose we have a unitary representation $U(\tau )$ of the full supersymmetric group in a Hilbert space
of supersymmetric functions

\begin{gather}
U(\tau )X(z)=X(\tau (z))
\end{gather}
A quantum supersymmetric field (free or interacting) and its vacuum are supposed to be supersymmetric invariant
such that this property is shared by the vacuum expectation values too. Denoting by $W_n (z_1 ,z_2 ,\ldots ,z_n )$
the n-point function (considered as superdistribution) we have simultaneous relations

\begin{gather}
Q_i W_n (z_1 ,z_2 ,\ldots ,z_n )=0, \quad \bar Q_i W_n (z_1 ,z_2 ,\ldots ,z_n )=0
\end{gather}
for $i=1,2,\ldots ,n$ where $Q_i ,\bar Q_i $ are the supersymmetric generators on the variables $z_i $. Because of
translation invariance $W_n $ depend on $x_i ,i=1,2,\ldots ,n $ through differences $x_i -x_{i+1} $. How does the
dependence of $W_n $ on $\theta_i ,\bar \theta_i $ look like? We start with the case $n=2$. Simultaneous
supersymmetric invariance imply for $W_2 (z_1 ,z_2 )$

\begin{gather}
(Q_1+Q_2)W_2 (z_1,z_2)=0 \\ (\bar Q_1+\bar Q_2)W_2 (z_1,z_2)=0
\end{gather}
where $Q_1,Q_2$ and $\bar Q_1,\bar Q_2$ act on the variable $z_1$ and $z_2$ respectively.\\ In order to solve
these equations we introduce new variables $\theta=\frac {1}{2}(\theta _1+\theta _2) $ and $\zeta =\theta
_1-\theta _2 $ together with their conjugates as well as (by translation invariance) the difference variable
$x=x_1-x_2 $ (for a similar argument see \cite{GS}). Note that by introducing the difference variable $x=x_1-x_2 $
the derivative $\partial _l $ in $Q_2,\bar Q_2 $ taken with respect to the second variable changes sign such that
in the new variables equations (9.2) take the form

\begin{gather}
(\frac {\partial }{\partial \theta ^{\alpha }}-i\sigma_{\alpha \dot \alpha }^l \bar \zeta^{\dot \alpha }\partial_l
)W_2=0   \\ (\frac {\partial }{\partial \bar \theta^{\dot \alpha }}-i\zeta^\alpha \sigma_{\alpha \dot \alpha }^l
\partial_l )W_2 =0
\end{gather}
where $W_2 $ is a function of $x=x_1-x_2 ,\theta ,\bar \theta , \zeta ,\bar \zeta $ . We want to solve this system
of partial differential equations in mixed commutative and non-commutative variables. A first (trivial and from
the physical point of view uninteresting) solution for $W_2 $ is a constant. Other solutions can be obtained in a
two step procedure by using in the first step the equation (9.3) to factorize from $W_2 $ the exponential $ exp
(i\theta \sigma^l\bar \zeta \partial_l )$ (for more details see also \cite{C2}). We write

\begin{equation}\nonumber
W_2 =exp (i\theta \sigma^l\bar \zeta \partial_l )D
\end{equation}
The first equation (9.3) implies $ \frac {\partial }{\partial \theta ^{\alpha }}D=0 ,\alpha =1,2 $ which means
that $ D $ is independent of $ \theta_{\alpha },\alpha =1,2 $. In the second step we write

\begin{equation}\nonumber
D=exp (-i\zeta \sigma^l \bar \theta \partial_l )E
\end{equation}
use (9.4), and conclude as above that $E$ is independent not only of $\theta $ but also independent of $ \bar
\theta $. These two exponentials cover the dependence of $W_2 $ from the variables $\theta $ and $\bar \theta $.
The residual dependence in $E$ is in $\zeta ,\bar \zeta $ and $x=x_1-x_2 $. Altogether the general solution of the
equations (9.3,9.4) in the $x,\theta ,\bar \theta ,\zeta ,\bar \zeta $-variables is (with the exception of the
constant solution) of the form

\begin{equation}
W_2 (x,\theta ,\bar \theta ,\zeta ,\bar \zeta )=exp[-i(\zeta \sigma^l\bar \theta -\theta \sigma^l\bar \zeta
)\partial_l ]E(x,\zeta ,\bar \zeta )
\end{equation}
where from invariance considerations $E$ is restricted to

\begin{gather}\nonumber
E(x,\zeta ,\bar \zeta )=E_1(x)+\zeta^2E_2(x)+\bar \zeta^2E_3(x)+ \\ +\zeta \sigma^l\bar \zeta
\partial_lE_4(x)+\zeta^2 \bar \zeta^2 E_5(x)
\end{gather}
with $E_i(x)=E_i(x_1-x_2),i=1 \ldots 5 $  Lorentz invariant functions (or distributions). The reality condition
would require real $E_i ,i=1,\ldots ,5 $ as well as $E_2 =E_3 $. Translating back to the $\theta ,\bar \theta
$-variables it is possible to relate (9.7) to the known invariants constructed with the help of the five invariant
operators $P_i ,i=c,a,T,+,- $ \cite{WB}.  \\ In the next section we will use this result in order to write down a
supersymmetric K\"allen-Lehmann representation for a scalar neutral supersymmetric quantum field theory. We
continue this section by extended our result from $n=2$ to general $n$. This is easily done by introducing the new
variables

\begin{gather}
\theta =\frac{1}{n}(\theta_1 +\ldots +\theta_n ),\quad \zeta_i =\theta_i -\theta_{i+1},i=1,2,\ldots ,n-1 \\ \bar
\theta =\frac{1}{n}(\bar \theta_1 +\ldots +\bar \theta_n ),\quad \bar \zeta_i =\bar \theta_i -\bar
\theta_{i+1},i=1,2,\ldots ,n-1
\end{gather}
such that the equations (9.5), (9.6) transform to

\begin{gather}
(\frac {\partial }{\partial \theta ^{\alpha }}-i\sum_{i=1}^{n-1}\sigma_{\alpha \dot \alpha }^{l_i} \bar
\zeta_i^{\dot \alpha }\partial_{l_i})W_n=0   \\ (\frac {\partial }{\partial \bar \theta^{\dot \alpha
}}-i\sum_{i=1}^{n-1}\zeta_i^\alpha \sigma_{\alpha \dot \alpha }^{l_i}
\partial_{l_i})W_n =0
\end{gather}
These equations can be solved in a two steps procedure exactly as in the previous case. The only difference is
that in the exponentials  in (9.11,9.12) we have to consider the corresponding sums over $\zeta_i ,\bar \zeta_i
,i=1,\ldots ,n-1 $. The result is

\begin{gather}\nonumber
W_n (z_1 ,z_2 ,\ldots ,z_n )=\exp [-i\sum_{i=1}^{n-1} (\zeta_i \sigma^{l_i}\bar \theta -\theta \sigma^{l_i}\bar
\zeta_i )\partial_{l_i}] \\ E(x_1-x_2 ,\ldots ,x_{n-1}-x_n ,\zeta_1 , \bar \zeta_1 ,\ldots ,\zeta_{n-1},\bar
\zeta_{n-1})
\end{gather}
The function E turns out as above, from invariance considerations, to be a finite sum of products of $ \theta_i
\theta_j , \bar \theta_i \bar \theta_j ,\theta_i \sigma^l \partial_l \bar \theta_j $ multiplied (applied) to
Lorentz invariant functions (distributions) depending on difference variables $ x_i -x_j $. The reality condition
imposes further obvious restrictions.

\section{Some other aspects of quantum superfields}

We come to the point of putting together the ingrediends and experience developed in the preceding sections in
order to indicate some other aspects of supersymmetric quantum fields which are accessible to our rigorous
methods. This can be done following ideas in the axiomatic quantum field theory \cite{SW}. Quantum fields are
defined as operator valued superdistributions. The supersymmetric invariance is formulated with the help of an
unitary representation of the super Poincare group on the postulated Hilbert space of supersymmetric functions.
Other axioms \cite{SW}, including positivity and locality (or weak locality), make no problems. As in the usual
case the reconstruction theorem based on the *-algebra of section 4 enables us to realize the supersymmetric
quantum fields as operator valued distributions on supersymmetric test function spaces starting from the n-point
functions. Before continuing on this line let us remark that our generalization is not trivial. It does not seem
to confirm the impression that in order to mathematically extend a theory to the supersymmetric case we only have
to adjust the prefix "super" on right places. Indeed, even the simple examples of the supersymmetric free field in
section 8 as well as the related generalized free field later on in this section show that among others, our
construction is centered on a non-trivial bona fide Hilbert space of supersymmetric functions and more important,
that this Hilbert space is intimately connected with an inherent Hilbert-Krein structure of the superspace
\cite{C1}. \\ We continue to restrict the consideration to the real massive vector field, which using another
terminology, could be called the supersymmetric scalar neutral field. As a first application we can mention the
supersymmetric K\"allen-Lehmann representation. Indeed the results of section 9  show that the most general
supersymmetric invariant two-point function is given as in (5.1) with the help of a kernel constructed with a
linear combinations of the invariant projections $P_i ,i=c,a,T,+,- $. An example is the standard kernel $P_c +P_a
-P_T $ in (5.1) but it turns out that this is not the most general one compatible with positivity. By computation
it can be proved that (in the massive case) the most general invariant and positive kernel is generated by
$\lambda_c P_c +\lambda_a P_a +\lambda P_+ +\bar \lambda P_- -\lambda_T P_T $ with $\lambda_c ,\lambda_a
,\lambda_T $ positive (or zero) and $ |\lambda ^2 | <\lambda_c \lambda_a $. This result allow us to introduce by
known formulas \cite{J} in analogy to (8.2-8.4) a whole family of (supersymmetric) quantum fields called
generalized free fields. We will not discuss them in details here.\\  Let us remind to the reader that from a
historical point of view the two main achievements of the axiomatic quantum field theory were the PCT and the spin
and statistic theorems \cite{SW,J}. We end this paper by formulating the PCT result in the supersymmetric
framework. First besides the $P,C,T$ transformations in quantum field theory (they do not act the Grassmann
variables; they act only the multiplet components) we introduce the $\Theta $ transformation by \cite{GGRS}

\begin{gather}\nonumber
\theta_{\alpha } \to i\bar \theta_{\dot \alpha }^{'} , \bar \theta_{\dot \alpha } \to i\theta_{\alpha}^{'} \\
\theta^{\alpha } \to i\bar \theta^{'\dot \alpha } , \bar \theta^{\dot \alpha } \to i\theta^{'\alpha}
\end{gather}
This is the PCT transformation of the spinor $\theta $ with components $\theta_{\alpha }$ \cite{SW}. The
supersymmetric PCT theorem \cite{SW} means the invariance of the theory to the transformation PCT supplemented by
$\Theta $ (it could be also called the $\Theta PCT $ theorem). The proof in our framework follows for example from
the general representation of the $n$-point functions in the scalar neutral theory given in (9.13). Indeed by
locality the space-time coefficients of the Grassmann variable products in the n-point functions (9.13) satisfy
PCT invariance. On the other hand, taking into account that $\theta_i \sigma^l \bar \theta_j
\partial_l  $ and $\theta_i \theta_j +\bar \theta_i \bar \theta_j $ are invariat under $\Theta $ it follows that
the supersymmetric n-point functions altogether are $\Theta PCT $ invariant.

\end{document}